# Earthquake and Geothermal Energy

**Surya Prakash Kapoor$^\$$**
60, Gagan Vihar Extension,
Delhi -110051 INDIA.

and

**Bhag Chand Chauhan$^\#$**
Department of Physics & Astronomical Science,
School of Physical & Material Sciences,
Central University of Himachal Pradesh, Dharamshala, Kangra -176215 INDIA.

**Abstract**

The origin of earthquake has long been recognized as resulting from strike-slip instability of plate tectonics along the fault lines. Several events of earthquake around the globe have happened which cannot be explained by this theory. In this work we investigated the earthquake data along with other observed facts like heat flow profiles etc…of the Indian subcontinent. In our studies we found a high-quality correlation between the earthquake events, seismic prone zones, heat flow regions and the geothermal hot springs. As a consequence, we proposed a hypothesis which can adequately explain all the earthquake events around the globe as well as the overall geo-dynamics. It is basically the geothermal power, which makes the plates to stand still, strike and slip over. The plates are merely a working solid while the driving force is the geothermal energy. The violent flow and enormous pressure of this power shake the earth along the plate boundaries and also triggers the intra-plate seismicity. In the light of the results reported by the California Energy Commission from the ongoing geothermal power project at the Big Geysers in California, we further propounded that by harnessing the surplus geothermal energy the intensity and risk of the impending earthquakes can be substantially reduced.

$ spkapoor@yahoo.com     # chauhan@iucaa.ernet.in

# 1. INTRODUCTION

Unlike other planets of the solar system earth is a highly dynamic. Some activity is relentlessly going on in the interior of it. Traversing a journey of hundreds of million years since its birth the earth has been going through a continuous geological change. As a result the continents on the surface had different shapes and were located in different positions from those we find them today [1]. The heat source inside the earth exerts pressure towards the surface where it leads to geo-dynamism and geothermal events like spectacular volcanoes, high heat flow regions etc…

The current understating of earthquakes is mainly based on the deformation of the plate tectonics along the fault lines. This is known as the theory of the Plate Tectonics. The rigid plates on the crust and lithosphere are moving slowly and continuously. Although most of the earthquakes have been found to occur along these fault lines associated with plate boundaries, yet there are certain events like mid-plate seismicity, peaceful (seamless) fault-lines and deep hypocenter etc… which are not adequately explained by this theory.

After a detailed investigation of the earthquake events and heat flow data etc…, mainly of the Indian subcontinent, in this paper we propose a hypothesis for the cause of earthquake which can successfully explain all of the above mentioned events and as well as the complete earthquake dynamics. According to this at the root it is the excess pressure of the geothermal power which sponsors the geo-dynamism, plate movements, and thereby earthquakes along the fault-line and thermally porous and brittle zone of the earth.

In the light of observed figures and facts we further argue that the plate movement and other seismic activity can be impeded by harnessing the surplus pressure amount of geothermal power as electricity etc…, at least in the hot springs pockets. As a result this intensity of the prospective earthquake can substantially be reduced. In the paper we have given a number of valid points and facts which associate the earthquake events more likely with the geothermal activity rather than just the plate movements.

The paper is organized as follows: Section 2 details the phenomenon of earthquake and the present understanding of its causes and predictions. In Section 3 we propose our hypothesis with its supporting arguments and observed facts by studying the relevant data of Indian subcontinent. However in Section 4 we discuss the potential benefits of harnessing the geothermal energy. The discussion and conclusions are summarized in Section 5.

## 2. EARTHQUAKE AND ITS CURRENT UNDERSTANDING

During the years 1912-15 a German meteorologist and geophysicist Alfred Wegener proposed that about 200 million years ago the earth existed as a supercontinent called *'Pangaea'*. Then it began breaking into smaller continents, which then drifted to their present positions. This drift was popularly known as *'Continental Drift'*. This idea was supported by evidences, like Fit to Continents, Fossil Evidence, Rock Type and Structural Similarities and Paleoclimatic Evidence etc... By 1968 the concept of continental drift united with another idea of *'Seafloor Spreading'* emerged as a more encompassing theory of *'Plate Tectonics'* [1]. The theory holds that the outer rigid lithosphere of earth is a spherical layer of about six major individual segments called *'Plates'*. These plates are Euro-Asiatic, African, Antarctic, Indo-Australian, American and Pacific, and several minor plates are positioned between them. The plates merged, not seamlessly, and the resulting lines across are known as *'Faults'*.

The earthquakes are caused by the friction on the boundaries of the plates moving together. This is considered to be the main reason that most of the earthquakes have been found to occur along these faults associated with plate boundaries. In the earthquake the vibration of earth is produced by a rapid release of energy. The energy is released as spherical wave fronts in all the direction starting from the source, the focus [1]. The source lies deep down inside the earth is called as hypocenter; however the corresponding point on the surface is known as epicenter. The waves so generated from the source are known as the seismic waves and the study of these waves is known as seismology.

The majority of earthquakes occur in the depths not exceeding tens of Kilometers. The earthquakes occurring at a depth of less than 70 Km are known as 'shallow-focus' earthquakes, while those with a focal-depth between 70 and 300 Km are commonly named as 'mid-focus' or 'intermediate-depth' earthquakes. However, 'the deep-focus' earthquakes may occur at much greater depths (ranging from 300 up to 700 Kilometers). These seismically active areas of subduction are known as Wadati-Benioff zones.

As per the theory of plate tectonics the earthquakes are mainly due to the strike-slip and stress-strain action of the plates along the fault lines. The rigid plates of the lithosphere are slowly, but nevertheless moving continuously. The lithosphere is a strong brittle layer overlying a weak ductile layer, which gives rise to two forms of deformation: brittle fracture, leading to earthquakes, in the upper layer, and aseismic ductile flow in the layer beneath. Although this view is correct, yet it is imprecise, and in ways that can lead to serious misunderstandings [2]. The author of [2] further illustrates that the earthquakes associated with strength and brittleness of the plates if taken much beyond the generality can lead to serious misinterpretations about earthquake mechanics.

There are certain events which the theory cannot explain clearly:

1) *The intra-plate seismicity and the hotspots far from plate edges*
2) *Earthquake followed by eruption and geothermal event [3]*
3) *Rare earthquake events along the portion of Gangetic plain on the plate boundary*
4) *Hypocenters are 700Kms beneath the lithosphere [4]*
5) *Earthquake is a fracture or strike-slip movement of plates but the observed event is a point-fracture, so called 'hypocenter/ epicenter'.*

In addition the current understanding is not capable enough to predict the potential earthquakes because of poor knowledge of earthquake dynamics. The mechanics of the rupture formation in the nucleation zone is not well understood. A number of methods have been developed for predicting the time and place in which earthquakes will occur, but the predictions cannot yet be made even to the level of a specific day or a month.

## 3. THE HYPOTHESIS

In the present study we mainly consider the earthquake data, seismic zones, heat flow profile and geothermal activity of the Indian geothermal province. In our investigation we found an encouraging correlation among all these indicators. As a result we propound that the gamut of earthquake phenomena—seismogenesis and seismic coupling, pre- and post-seismic phenomena, and the relatively weak dependence of earthquakes on slip and stick motion along the fault lines—all are the manifestations of the geothermal power. It is primarily this energy, which fuels the stability and movement of the plates and after penetrating, piercing, puncturing and punching through the surface of earth it emerges out and enters into open space through the oceanic spreading centers, subduction zones, plate collision areas, volcanoes, hotspots and hot springs etc. The violent flow and enormous pressure of this power may result in hazardous earthquakes along the plate boundaries and even at the mid part of plates.

Historically the earth was formed as a fire ball resulted by the collisions of millions of burning meteoroids originated from the cosmic debris. Since its inception the fire ball is still burning from inside and to a good extent the heat is shielded by the outer layers. A global terrestrial heat flow of 44.2 Terawatts (TWts) power is replenished by the self sustained natural nuclear fission reactor at a rate of 30 TWts. The energy widely dispersed in the remaining body of the earth energizes the solid earth, including core and mantle convection, plate motion, mountain building, earthquakes and volcanism. The earth's internal heat powers all geodynamic processes along with generation of geomagnetic field. It has been estimated that $0.3 \times 10^7$ M Wt ($1.1 \times 10^{14}$ Btu / hr) of heat is being dissipated by the earth into atmosphere continually and the average global heat flow value is 87 milli Wts / $m^2$. It may be noted that the high heat flow values occur on the plate boundaries. Besides the plate margins the high heat flow values occur in areas under tension where the crust is thinner than normal.

The high heat flow values on the continents are from volcanic and tectonically active regions while the highest values in the oceans are found near the area of oceanic ridges. In other words, we can say that on the surface of the earth the phenomena of geothermic and its off-shoot geodynamics go hand in hand with each other at plate boundaries. As per surface observations and hotspot data the high temperature (220 °C - 350 °C) geothermal resources are located at plate boundaries and low to intermediate temperature (50° C - 220 °C) resources are spread in the midst of plates.

We argue that the high temperature geothermal resources create earthquakes at plate boundaries whereas low to intermediate temperature geothermal resources fund intra-plate seismicity. Most of the geothermal energy inside the earth escapes as heat and eventually radiates in outer space through the volcano, hotspots and hot springs etc…. However a tiny bit of this energy is released in earthquakes, where the passage is rather tough.

According to a latest study the release of mud could have been a natural response of an earthquake after all [3]. This argument also favors the geothermal origin of the earthquakes. In the interior the excess pressure of geothermal power fractures the weak part of the plate boundary and earthquake takes place and the mud starts coming out through the brittle part of the plate, if any. The mud flow or eruption may be quite far from the epicenter. Most of the times it just fractures the plate and no mud comes out because of no easy channel available. So earthquake is there but no mud eruption is recorded.

**3.1 Indian Earthquake Zones and Heat Flow Regions**
As per the studies of the Structural Civil Engineers [5] "The intuitive development of the first seismic zone map and of the earthquake resistant features for masonry buildings took place in 1930's and formal teaching and research in earthquake engineering started in late 1950's. The first code of Practice of Bureau of Indian Standards on Earthquake Resistant Design of Structures was published in 1962 (IS 1893-1962). The seismic zoning of the country was brought out in the Code which demarcated a major part of peninsular India as non-seismic. The IS code has undergone several revisions and the latest code (IS 1893-2002) has been divided into five parts, each applicable for different types of structures. The magnitude or intensity of expected earthquake in these regions is based on past data spanning a few hundred years and is not adequately understood. It is generally known that the Himalayan region has the highest seismic hazard in the country. The Kashmir earthquake (2005) for example, was more intense than the specifications of IS codes for that region. In regions away from the Himalayas, the seismic hazard is even more poorly understood. The Killari earthquake (1993) occurred in a region categorized as seismic Zone-I, which implied low probability of damaging earthquakes. There is an urgent necessity to improve our understanding of the seismic hazard in the country, so that the structures are designed to consider the appropriate intensity of earthquake shaking". The seismic zone map of India is shown in the map below:

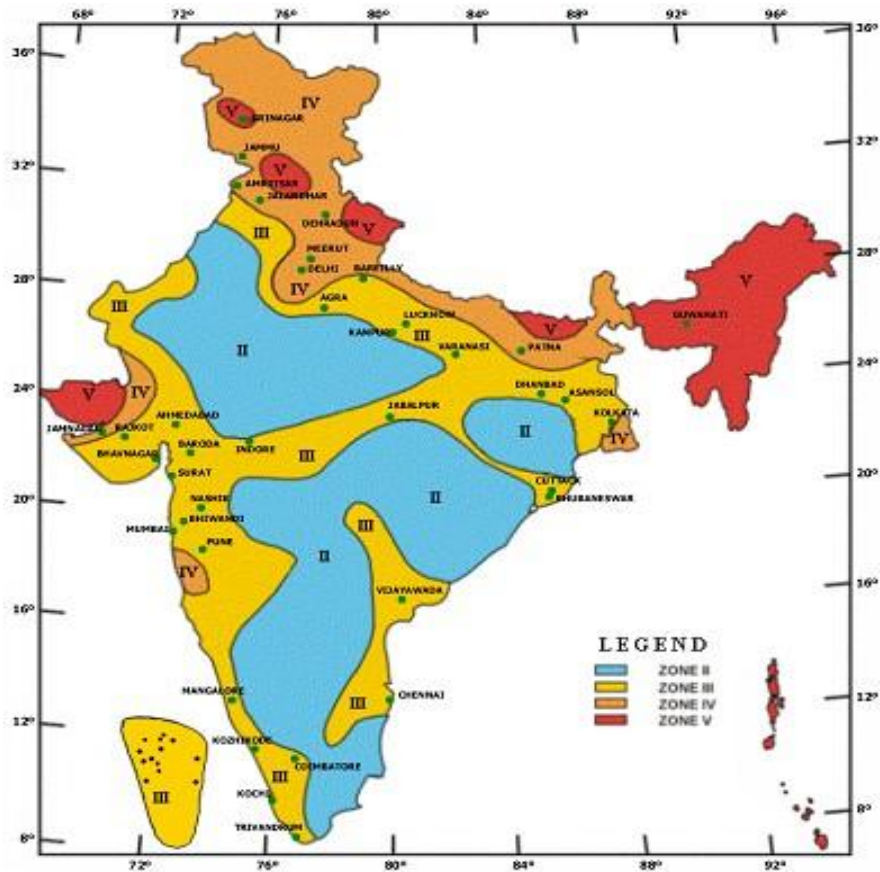

*Figure 1: The seismic zone map of India [5]*

The region is divided into four (II-V) Zones of various seismic prone areas. The Zone-V corresponds to the highest scale of earthquake and the Zone II shows the region of the lowest seismic probability. While comparing it with the Heat Flow Map [6] given below in Figure 2, one can easily validate the fact that the highest heat flow (Zone-I) corroborates with the highest seismic prone area (Zone-V) and the lowest heat flow region (Zone-V), to a good extent, overlaps with the lowest seismicity area (Zone-II). It can be concluded that the amount of surplus geothermal energy accumulated and exploded in the form of earthquake is directly proportional to heat flow value of the associated geothermal area. The red regions (Zone-I) having highest heat flow value more than 180 milli Wts/m$^2$ spawn the strongest earthquakes most frequently. The Himalayan geothermal area has highest heat flow value 498 milli Wts / m$^2$ accounting for almost 60% share of the aggregate of devastating earthquakes in India.

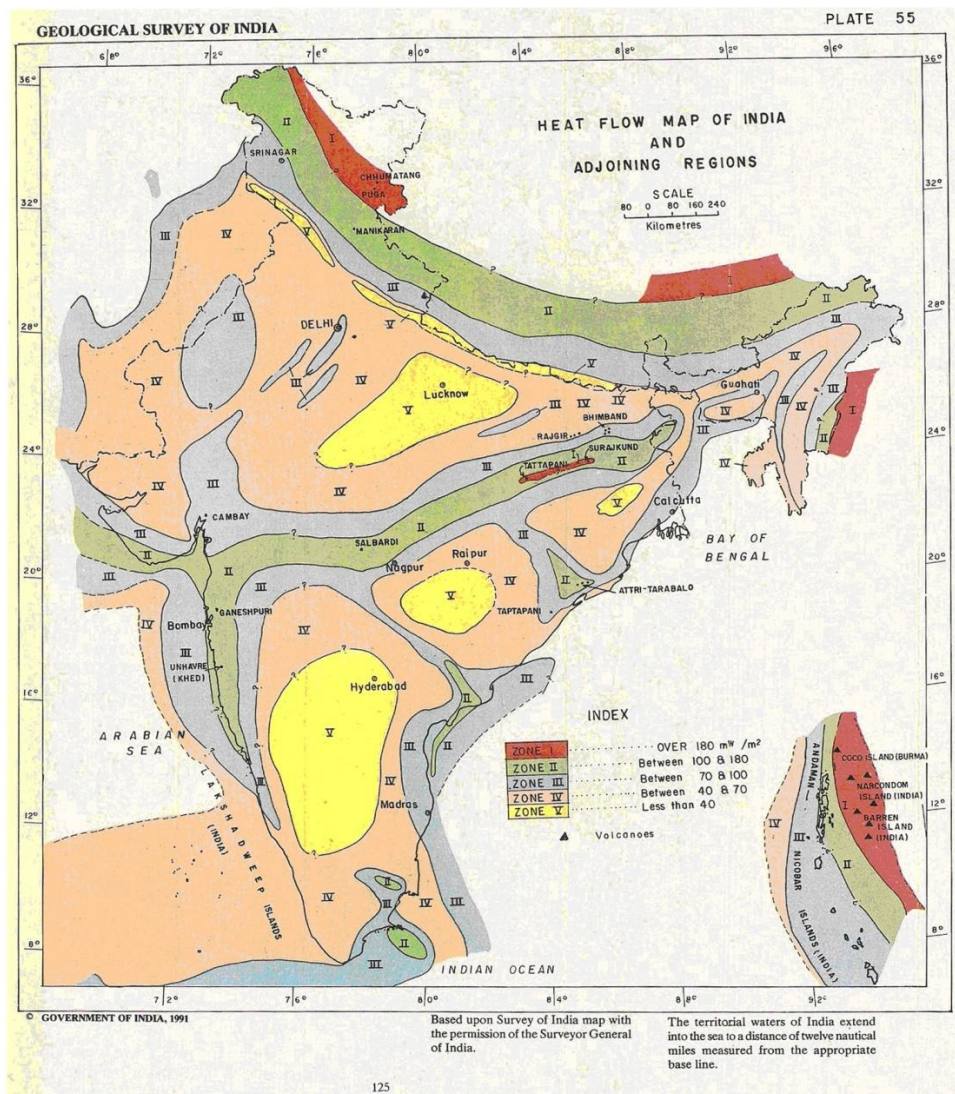

*Figure 2: Heat flow map of India [6].*

The Heat Flow Map clearly shows the cities like Hyderabad, Bangalore, Lucknow are falling in yellow regions (Zone-V) having heat flow value less than 40 milli Wts/m$^2$ the regions are earthquake proof because the heat flow value is so low that it cannot sponsor earthquake. The portion of Gangetic Plain comprising of Kanpur, Allahabad and Varanasi situated in the lowest heat flow Zone-V and although in the vicinity of Himalayan plate boundary, yet has not witnessed any earthquake hazard as per record.

It may be noted that unlike the heat flow map (Figure 2) the seismic zone map (Figure 1) is not based on any concrete scientific foundation rather it is more a hit and trial basis map. We believe that the overlap of the respective zones of Figure 1 and 2 will be more once we get the accurate one.

**3.2 Indian Geothermal Provinces**

The geothermal hotspot regions in India are divided into seven provinces namely Himalayas, Sohana, Cambay, West Coast, SONATA, Godavari, and Mahanadi. These seven geothermal provinces are characterized by high heat flow value (78-468 milliWt/m$^2$) and thermal gradients (47-100 $^0$C/Km) and discharge about 400 thermal springs within Indian jurisdiction. After the oil crisis in 1970s, the Geological Survey of India conducted reconnoiters on them in collaboration with UN organization and reported the results in several of their records and special publications [6]. The investigations have identified several sites which are suitable for power generations as well as for direct use. These provinces are capable of generating 10,600 MW of power [7].

The first pilot binary 5 KWt power plant using R 113 binary fluid was successfully operated by the Geological Survey of India at Manikaran which proved the power producing capability of this province. The space heating experiments were also successfully conducted using thermal discharge by the Geological Survey of India. The seven geothermal provinces and the Himalaya province are shown in the picture (Figure-3) below.

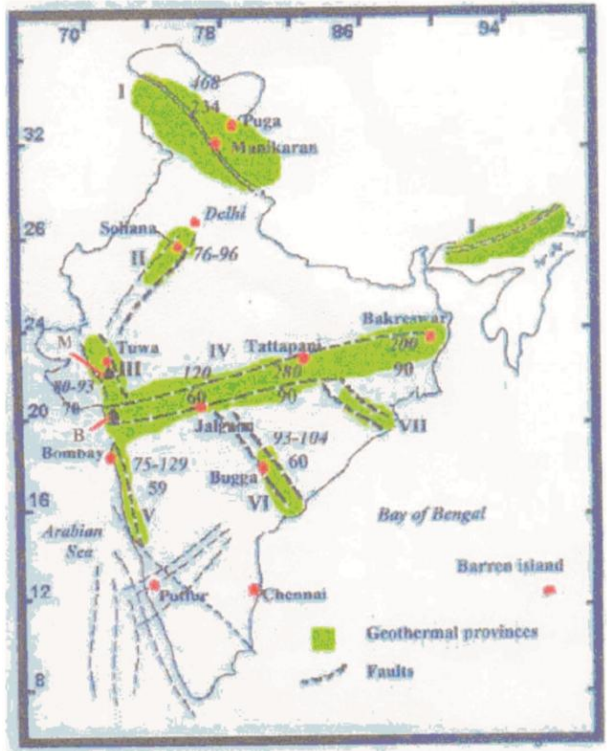
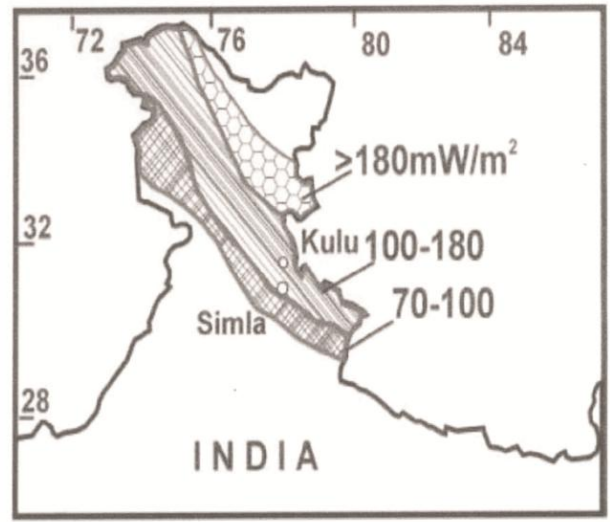

*Figure 3: Geothermal provinces and the Himalayan province map of Indian subcontinent [7]*

### 3.3 Indian Geothermal Provinces and Earthquakes

The collision of the Indian plate with the Eurasian plate resulted in the formation of the Himalayas about 45 million years ago and as one of the largest geothermal belt, over 150 Km wide extends 3000 Km through parts of India, Tibet, Yunnan (China), Myanmar and Thailand replete with more than 1,000 hot spring areas. Over 150 of these areas are hot enough to generate electricity. The arc of volcanic islands incorporating the Coco Island (Burma), Nar-Condom Island (India), and Barren Island (India) is situated on the eastern tail end of Himalayan Geothermal Belt.

The year-wise data of the major earthquakes of Richter scale (>4) in the peninsula and the associated geothermal provinces have been tabulated below since 1819-2007:

| Year | Region | Magnitude | Toll | Name of Associated Geothermal Province |
|------|--------|-----------|------|----------------------------------------|
| 1819 | Kutch, Gujrat | 8.0 | 2,000 | Cambay |
| 1885 | Sopore, JK | 7.0 | 2,000 | Himalaya |
| 1897 | Shilong | 8.7 | 1,542 | Himalaya |
| 1905 | Kangra, HP | 8.0 | 19,500 | Himalaya |
| 1918 | Assam | 7.6 | NA | Himalaya |
| 1930 | Assam | 7.1 | NA | Himalaya |
| 1934 | Bihar – Nepal | 8.3 | 10,700 | Himalaya |
| 1941 | Andaman Island | 8.1 | NA | Himalaya |
| 1943 | Assam | 7.2 | NA | Himalaya |
| 1950 | Arunachal | 8.5 | 1,526 | Himalaya |
| 1956 | Gujarat | 7.0 | 113 | Cambay |
| 1960 | Delhi | 6 | Nill | Sohana |
| 1967 | Koyna, Maha | 6.5 | 177 | West Coast |
| 1970 | Bhadrachalam, AP | 6.5 | Not Known | Godavari |
| 1970 | Broach, Gujrat | 5.7 | Not Known | Cambay |
| 1975 | Himachal Pradesh | 6.5 | Not Known | Himalaya |
| 1988 | Bihar – Nepal | 6.4 | 900 | Himalaya |
| 1991 | Uttarkashi, UP | 6.6 | 2,000 | Himalaya |
| 1993 | Latur, Maha | 6.3 | 9,748 | West Coast |
| 1997 | Jabalpur, MP | 6.0 | 38 | SONATA |
| 1999 | Chamoli, UP | 6.8 | 100 | Himalaya |
| 2001 | Bhuj, Gujarat | 8.7 | 19,988 | Cambay |
| 2004 | Andaman Island | 7.5 | 2,000 | Himalaya |
| 2005 | Muzafarabad, JK | 8.5 | 36,000 | Himalaya |
| 2007 | Bahadurgarh, HR | 4.3 | Nil | Sohana |

*Table 1: Earthquakes sponsored by the geothermal provinces of India*

From the above table it is quite evident that the epicenters of all the big damaging earthquakes of India during last 200 years are strictly located in the jurisdiction of above mentioned seven geothermal provinces of India. This is a glaring fact in support of our hypothesis that earthquakes are the emission of surplus unmanaged geothermal energy accumulated in the respective geothermal provinces during last two centuries. On the other hand, the portion of Gangetic Plain comprising of Kanpur, Allahabad and Varanasi, although situated in the vicinity of Himalayan plate boundary, but lying in non-geothermal province (a low heat flow region), has not witnessed any earthquake hazard in the last 3000 years. This shows failure of the influence of mechanical energy accumulation parameters like strike-slip and stress-strain at plate boundary / hydro-seismicity whatsoever and success of geothermal energy accumulation parameters like heat flow value at plate boundary.

## 3.4 Intra-Plate Seismicity

There are certain earthquake events which occurred quite far from the plate boundaries and fault lines known as intra-plate seismicity. As per the theory such events are attributed to reactivation along the old faults called intra-plate faults. However, there are a number of arguments to defend, but are quite hard to swallow. On the basis of the observed facts we propose the geothermal origin of the intra-plate quakes.

During 2000BC the Indus Valley civilization was spread over in Sapta Sindhu (Saraswati, Indus, Ravi, Beas, Sutlej, Jhelam, Chenab) area of North West India. The recovery of melted and burnt remains from the site during excavation by archeologist stands testimony to the unique feature that this civilization was devastated by severe earthquakes [8]. Examples of the earthquake effects are present in Banbhore in the Indus Delta, Brahmanabad, and the Harappa sites of Kalibangan and Dholavira. As per our hypothesis those earthquakes were triggered by a moderate magmatic extrusion resulting from excess geothermal energy pressure accumulated beneath Sohana geothermal province. This was a blatant case of intra-plate seismicity in ancient India.

A swarm of mild earthquakes have been rattling the North West India comprising of Delhi, Haryana, Rajasthan and Punjab. The frequency of micro-seismicity has increased alarmingly. For the past few years, an 'earthquake swarm' has been reported from the southern Punjab basin with its epicenter near the town of Jind of Haryana. Apart from Jind, tremors were also felt in nearby areas of Kaithal, Narnaul and Hissar districts: rattling of doors, windows, shop shutters, shaking of furniture and cracks in the walls have been reported. The recent earthquake dated 26[th] November, 2007 having epicenter at Bahadurgarh (Delhi, Haryana Border) measuring 4.3 degree on Richter scale has sufficiently scared Delhiites and pricked their vanity. This was the fourth consecutive earthquake in year 2007 having epicenter at Bahadurgarh; again an event of mid-plate seismicity.

In addition to this the earthquakes of very high magnitude reported in the Western and South Indian districts like Bhuj, Jabalpur, Latur & Bhadrachalam are vivid examples of the intra-plate seismicity and geothermal origin of these events.

### 3.5 Hypocenter: The Focus of Earthquake

According to the observations the earthquake begins by an initial rupture at a point on the fault surface, a process known as nucleation. The scale of the nucleation zone is not clear, but some evidence shows that the dimensions are around 100 meters. The possibility that the nucleation involves some sort of preparation process is supported by the observation that about 40% of earthquakes are preceded by foreshocks. Once the rupture has initiated it begins to propagate along the fault surface.

The fault is a line where the plates are fused and along which the plates slips and strikes. The questions arise: 1) Why the earthquake origin is identified as a point (focus) i.e. 'hypocenter /epicenter'?  2) Why such a huge plate of thousands of kilometers could slip or rupture just at a point kind of thing i.e. 100 meters length? In principle, if there is a slip along, it should be a line type extended along the entire fault line and the seismic waves should propagate as the cylindrical wavefronts rather than the spherical one. As a result of which the several places over the distance of thousand kilometers should have experienced the quake of same magnitude on the Richter scale. In general, no evidence of this kind of quake shows that it cannot be a slip of the plates at all.

As per the nature of quakes observed so far i.e. point as the source clearly suggest the geothermal origin of the earthquakes even along the plate boundaries. The excess heat and pressure of geothermal energy puncture and produce fractures along the weak fault lines and other delicate areas, even on the mid of plates. We can conclude the phenomenon of earthquake is just like a puncturing of football; but not exactly.

### 4. HARNESSING GEOTHERMAL ENERGY

Geothermal energy is the immense store of heat (~$10^{13}$ EJ) in the earth, which alone would take over $10^9$ years to exhaust. So the geothermal source is an extremely large and self sustained natural gift [9]. These power rates are more than double humanity's current energy consumption from all primary sources, most of which are not recoverable. As per the recent report of the International Geothermal Association (IGA) 10,715 Megawatts (MWts) of geothermal power in 24 countries is online and out of which about 67,246 GWts h of electricity is being generated.

The science and technology of geothermic is progressing. USA is the world's largest producer of geothermal electricity. The first geothermal plant, opened at 'The Big Geysers' in California in 1960 continues to operate successfully. The California Energy Commission (CEC) is running this largest geothermal energy plant of 1800 MWts. We are citing the encouraging results at page 19 of the CEC report [18]: *"If one considers the energy release over time, rather than just a count of earthquakes, then one gets a different picture. We took the same information in Figure 4 from 1984 to the present and looked at the energy release over time by using an energy - magnitude relation Log 10E = 11.4 + 1.5 M, (E = energy in ergs, M = magnitude of the event). Figure 5 shows the rate of seismicity (total events above M = 1.5) for the Geysers area since 1984, If one converts the magnitudes to energy one obtains the results in Figure 6, As can be seen the rate of energy release is actually decreasing as a function of time"*

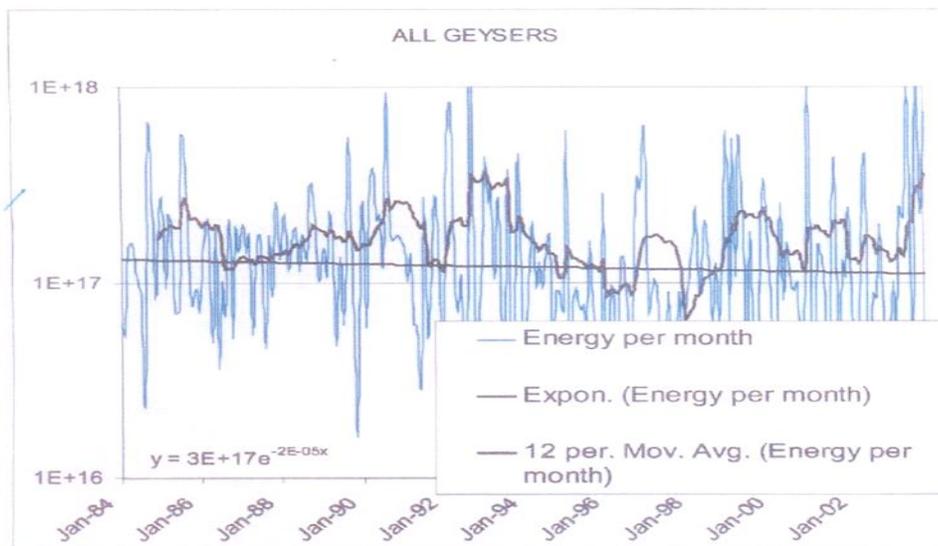

Figure 6. Energy release over time since 1984, the line is a linear fit to a 12-month moving average.

*Figure 4: Energy release through the geothermal the Geysers: Figure 6 of the Report [10]*

The decreasing rate of energy release in the above figure reflects optimism about the high magnitude earthquake hazard free future of the Geysers Area. Harnessing geothermal energy is multifold rewarding. Besides fulfilling the electricity requirement of 10,00,000 house hold units of North California in an Eco friendly manner; page 20 of the report further reads. *"To date there has been no faults mapped in The Geysers which would generate a magnitude of 5.0 or greater. This is not an absolute guarantee that one would not happen, but does lower the likelihood".*

It has been found that the geothermal field development and expansion has resulted in seismic activity, though many of these induced micro-earthquakes require sensitive instrumentation to be detected. The Environment Impact Report (EIR) determined that a geothermal facility would induce less than significant increases in seismic activity [11].

The largest earthquake ever detected in The Geysers area measured 4.6 on the Richter scale; while seismic activity elsewhere in the region can be much more dangerous. In 1969, Santa Rosa, California, 40 miles from the geothermal site, experienced an earthquake of magnitude 5.7 in 1969. The experts do not treat the seismic activity at The Geysers as a significant concern compared to the larger-magnitude seismicity in the region, and therefore put specifically no focus on monitoring efforts in the Geysers field. The project located in Basel, Switzerland, was also deemed too dangerous and capable of triggering earthquakes. The project has been on hold since 2006, when it caused a 3.4 magnitude earthquake, thousands of aftershocks and millions worth of damage in Basel, a town of about 167,000 people.

It may be noted that both the projects are based on an 'enhanced geothermal system', which fractures bedrock by blasting and high pressure water is circulated through the cracks to heat it and produce steam that powered the turbines of a power plant. It is not surprising that the fracturing process of the 'enhanced geothermal system' and introducing high pressure water into the bores can cause earthquakes, because the water entering into the cracks produce steam of very high pressure and do not find an immediate exit except the one used for power generation. The excess steam enters into the nearby area through cracks and cause earthquake. In addition, naturally, both projects were also based in areas with a history of seismic activity. So rather discouraging the harnessing of geothermal energy there is a need to continue the R &D and stay most likely with the natural hot-springs area and wait for the next technological advancement in the 'enhanced geothermal system'.

As argued before, we again conclude that it is the surplus geothermal energy which funds the movement of the plate tectonics; results point fracture along the active fault and intra-plate regions and after penetrating, piercing, puncturing and punching through the surface of earth emerges out and enters into open space. So, it is mainly this power which results in hazardous earthquakes at the oceanic spreading centers, subduction zones, and plate collision areas and even in the midst of plates. In the light of the above results from the Big Geysers if we harness this excess heat flow pressure through the active fault lines or geothermal provinces, we can minimize the risk and hazard of the impending earthquakes and moreover, we get electricity as a bonus.

## 5. DISCUSSION AND CONCLUSIONS

About 500,000 earthquakes occur each year which are detectable with current instrumentation and around 100,000 of these can be felt. Minor earthquakes occur nearly constantly around the globe and Major earthquakes occur less frequently, but in the identified pockets. The actual cause of the continental drift i.e. motion of the plates is not clear yet, it is however, interesting that the tectonic activity—the volcanic eruptions---has also been discovered in some other planets and satellites. This shows a similar kind of processes going on in interior of those bodies.

The plate tectonics theory fails to explain vividly a wide range of earthquake events as discussed in the section 2. In our investigation we found a quite convincing overlaps in the earthquake events, seismic zones, heat flow profile and geothermal provinces in the Indian peninsula. We reached to a conclusion and proposed a hypothesis that in the root it is the geothermal energy which gives rise to plate movements, inter-plate and the intra-plate earthquake events. The excess pressure of the geothermal power punctures and pierces the physically weaker section along the fault line and even mid part of the plates.

We support our hypothesis with the facts and figures of the data and earthquake events discussed in detail above logically. In addition to that the deep-focus earthquakes having hypocenter as deep as 700 Kms are questionable whereas the thickness of lithosphere plates is merely 100 Kms. The occurrence of such earthquakes cannot be explained in terms of release of accumulation of stress and strain along the plate boundaries of lithosphere. These earthquakes cannot be dubbed as mechanical energy release events. Obviously, these volatile explosions on account of melting & boiling of subduction plate fragments are geothermal energy release events.

The results and the inferences drawn in this work are quite general and not only pertaining to the Indian subcontinent. As the source of geothermal energy generation is spherically symmetric and the structure of the upper mantle layer and lithosphere is quite similar throughout the globe so the conclusions are equally applicable to the entire geo-dynamics subject to the nature of fault lines.

It is thought to be virtually impossible to control the course of natural events, but on the basis of our findings and the California Geysers report we are convinced to put forward such a bold statement that harnessing the geothermal energy as house heating and electricity generation can reduce the impact of earthquakes / tsunamis / volcanism. So, in order to cope with the escalating energy crisis and as the measure of earthquake disaster management there is dire need of harnessing the surplus pressure amount of geothermal power as electricity etc…, at least in selected pockets of natural geothermal provinces. It is based on sound principles of energy engineering in contrast to the theory of plate tectonics, which incorporates assumptions of significant value. To avoid the risk of earthquakes during harnessing power through the 'enhanced geothermal systems' we will still have to wait for a higher technology.

**Acknowledgement**

One of the authors B. C. Chauhan is thankful to the Inter University Centre for Astronomy & Astrophysics (IUCAA) for providing necessary facilities during the completion of this work.